\documentclass[referee]{raa}            % referee version: for submission
\usepackage{graphicx,times}  
\graphicspath{{Pic/}}
\usepackage{graphicx,times}             %for PS/EPS graphics inclusion, new
\usepackage{natbib}
\usepackage{lineno}
\usepackage{amssymb,amsmath}
\bibpunct{(}{)}{;}{a}{}{,}

\usepackage{listings}
\definecolor{codegreen}{rgb}{0,0.6,0}
\definecolor{codegray}{rgb}{0.5,0.5,0.5}
\definecolor{codepurple}{rgb}{0.58,0,0.82}
\definecolor{backcolour}{rgb}{0.95,0.95,0.92}
\lstdefinestyle{mystyle}{
    backgroundcolor=\color{backcolour},   
    commentstyle=\color{codegreen},
    keywordstyle=\color{black},
    numberstyle=\tiny\color{black},
    stringstyle=\color{codegreen},
    basicstyle=\ttfamily\footnotesize,
    breakatwhitespace=false,         
    breaklines=true,                 
    captionpos=b,                    
    keepspaces=true,                                   
    numbersep=5pt,                  
    showspaces=false,                
    showstringspaces=false,
    showtabs=false,                  
    tabsize=2,
    language=Python
}
\usepackage[pagebackref=true]{hyperref}

\begin{document}
%\linenumbers

  \title{In-Orbit GRB Identification Using LLM-based model for the CXPD CubeSat}
%   \subtitle{I. Place Your Subtitle Here}

   \volnopage{Vol.0 (20xx) No.0, 000--000}      %%preserved for Editor. DOn't remove!
   \setcounter{page}{1}          %%starting page, preserved for Editor. DOn't remove!

\author{Cunshi Wang \inst{1,2} \thanks{\url{wangcunshi@ucas.ac.cn}}
   \and Zuke Feng \inst{3} \thanks{Co-first author}
   \and Difan Yi \inst{4}
   \and Yuyang Li \inst{1,2}
   \and Lirong Xie \inst{3}
   \and Huanbo Feng \inst{3}
   \and Yi Liu \inst{5}
   \and Qian Liu \inst{4}
    \and Yang Huang \inst{1,2} 
    \and Hongbang Liu \inst{3}
    \and Xinyu Qi \inst{5}
    \and Yangheng Zheng \inst{4}
    \and Ali Luo \inst{2}
   \and Guirong Xue \inst{5} \thanks{\url{grxue@tianrang-inc.com}}
   \and Jifeng Liu\inst{1,2,6}
   }

   \institute{    
    School of Astronomy and Space Sciences, University of Chinese Academy of Sciences, Beijing 100049, People's Republic of China \label{UCAS}
    \and
   National Astronomical Observatories, Chinese Academy of Sciences,
    20A Datun Road, Chaoyang District, Beijing 100101, People's Republic of China \label{NAOC}
    \and 
    Guangxi Key Laboratory for Relativistic Astrophysics, School of Physical Science and Technology, Guangxi University, Nanning 530004, People's Republic of China \label{GXU}
    \and
    School of Physical Science, University of Chinese Academy of Sciences, Beijing, 100049, China
    \and
    Zhejiang Lab, Wenyi West Rd.2880, Yuhang, Hangzhou, People's Republic of China \label{ZJL}
    \and
    New Cornerstone Science Laboratory, National Astronomical Observatories, Chinese Academy of Sciences, Beijing 100101, People's Republic of China \label{NCS}
\vs \no
{\small Received 20xx month day; accepted 20xx month day}
}

\abstract{
To validate key technologies for wide field-of-view (FOV) X-ray polarization measurements, the Cosmic X-ray Polarization Detector (CXPD) CubeSat series has been developed as a prototype platform for the Low-Energy X-ray Polarization Detector (LPD) onboard the POLAR-2 mission. The wide-FOV design significantly increases the complexity of the background environment, posing notable challenges for real-time gamma-ray burst (GRB) identification. In this work, we propose an in-orbit GRB identification method based on machine learning, using simulated spectral data as input. A training dataset was constructed using a Geant4-based simulator, incorporating in-orbit background and GRB events modeled within the 2–10 keV energy range. To meet the computational constraints of onboard processing, we employ a multimodal large language model (MLLM), which is fine-tuned using low-rank adaptation (LoRA) based on miniCPM-V2.6 and quantized to 4-bit precision. The model achieves perfect classification accuracy on validation data and demonstrates strong regression performance in estimating GRB spectral indices, with an RMSE of 0.118. Furthermore, we validate the feasibility of onboard deployment through a simulated satellite data processing pipeline, highlighting the potential of our approach to enable future real-time GRB detection and spectral analysis in orbit.
}

   \authorrunning{Cunshi Wang and Zuke Feng et al}            %author_head in even pages
   \titlerunning{Photometry of $\delta$ Sct and Related Stars (I) }  % title_head in odd pages

   \maketitle
%% The author head (on even pages) and the title head 580 (on odd pages) will be
%% automatically extracted from \author{} and \title{}. Whenever the title is too long,
%% you will be asked to supply a shorter one by inserting either \authorrunning{} or
%% \titlerunning{} before \maketitle. Anyway, you can specify your own heads.
%%
%%
%% Note: In the following text body of your manuscript, please note several differences from
%%       other major journals:
%% (1) \subsection{Please Capitalize the First Letter of Each Notional Word in Subsection Title}
%% (2) Please Capitalize the First Letter of Each Notional Word in all tables' captions

%
%________________________________________________ sections below
%
\section{Introduction}           %% first-level sections will be auto-capitalized
\label{sect:intro}

Gamma-ray bursts (GRBs) are among the most luminous and transient astrophysical phenomena, capable of releasing an extraordinary amount of energy on timescales of seconds. Despite decades of observational efforts, the fundamental physics governing GRBs—such as jet composition, magnetic field geometry, and radiation mechanisms—remains poorly constrained \citep{galaxies9040082}. Polarization measurements offer a complementary probe to traditional spectroscopy, and their joint analysis continues to provide critical insights into these unresolved questions \citep{kumar2015physics,10.1111/j.1745-3933.2009.00624.x,10.1093/mnras/sts219,10.1111/j.1365-2966.2010.17600.x,Zhang_2002,Bégué_2016,Rees_1994}.

Recent advances in gamma-ray polarimetry have revealed a more nuanced picture of GRB prompt emission than previously expected. The POLAR mission, for instance, reported surprisingly low polarization degrees (PDs) — around 10\% on average — for a sample of bursts \citep{zhang2019detailed}. Rather than indicating the absence of polarization, these modest values likely reflect dynamic changes in the polarization angle (PA) within individual pulses, effectively diluting the net signal. This finding stands at odds with traditional synchrotron models, which often predict significantly higher PDs under ordered magnetic field scenarios \citep{granot2003linear}, and also contrasts with earlier reports of highly polarized bursts \citep{guan2023interpreting}. The resulting discrepancy paints a more intricate portrait of GRB emission physics, one that calls for sharper observational tools and more flexible theoretical approaches.

POLAR-2, the planned successor to the POLAR mission, is designed to probe the polarization properties of GRB prompt emission with improved sensitivity and a broader energy coverage. This next-generation instrument aims to address open questions regarding the energy dependence of polarization signals during the prompt radiative phase. Theoretical models suggest that PD can vary significantly with photon energy: synchrotron and Compton drag scenarios generally predict lower PDs at lower energies \citep{toma2009statistical}, while photospheric emission models may yield stronger polarization signals in the few-keV range \citep{lundman2018polarization}. By capturing this energy-dependent behavior in the prompt emission, POLAR-2 has the potential to discriminate among competing emission mechanisms and constrain the physical conditions in GRB jets.

To access the largely unexplored soft X-ray polarization regime, POLAR-2 incorporates a dedicated Low-Energy X-ray Polarization Detector (LPD), designed to operate in the 2–10 keV band using an array of photoelectric polarimeters. In contrast to missions such as IXPE and eXTP \citep{WEISSKOPF20161179, zhang2019enhanced}, which employ similar detection principles but depend on focusing optics for deep observations of persistent sources,  the LPD adopts a novel wide field of View (FOV) configuration optimized for survey-mode observations of transient events \citep{feng2025polarization}. This approach enables  the real-time detection of GRB prompt emission and other short-lived phenomena across the soft X-ray sky.

The Cosmic X-ray Polarization Detector (CXPD) CubeSat series serves as a technology demonstrator for the LPD, aiming to validate critical components required for wide FOV X-ray polarization measurements. While the wide FOV design is essential for capturing transient events such as GRBs, it also introduces a higher rate and diversity of background events. To identify transient sources amid large volumes of background data and effectively reduce the satellite’s data downlink burden, we propose an in-orbit GRB identification strategy for CXPD based on a multimodal large language model (MLLM), which ingests simulated spectral data to accurately distinguish prompt GRB signals from complex background noise.

Large language models (LLMs) can offer significant flexibility by representing data in free-form text. They eliminate the need for fixed-dimensional tensor inputs, thereby supporting dynamic input spaces and enabling multi-task learning through model scaling. Moreover, fine-tuning allows rapid adaptation to novel tasks, and emerging techniques now enable LLMs to quantify predictive uncertainty—offering a powerful new paradigm for general regression problems \citep{song2025omnipredlanguagemodelsuniversal, liu2024autotimesautoregressivetimeseries}.

The application of artificial intelligence in astronomy has evolved from traditional machine learning methods toward large-scale, transformer-based architectures. Recent studies have shown that—when equipped with appropriate data encoding strategies—these models can effectively perform core astronomical data processing tasks previously thought to require highly specialized, domain-specific pipelines. \citet{2025liLLM} applied several LLM-based classifiers to Kepler light curves and achieved over 95\% accuracy in distinguishing among five classes of variable stars. \citet{Du_2025deepAP} developed an adaptive photometry pipeline where a Vision Transformer (ViT) first assesses whether aperture photometry is feasible from an image, followed by automatic aperture selection via a sequence of deep models. \citet{Sun_2025DDPM} employed a Denoising Diffusion Probabilistic Model to reconstruct weak spectral lines in low signal-to-noise ( S/N $ \sim$ 10  –15) LAMOST spectra in the  $ g $ -band. Most recently, \citet{Zhao_2025DL} leveraged Low-Rank Adaptation (LoRA) to fine-tune multimodal LLMs—specifically DeepSeek-VL-7B and InternVL2-40B—on visual question answering (VQA) tasks involving radio astronomy images, achieving a remarkable 98.9\% accuracy, substantially outperforming both conventional deep learning models and off-the-shelf multimodal LLMs.

The application of MLLMs to in-orbit computation aboard satellites offers two key advantages. First, the data downlink capacity of satellites is constrained by the limitations of communication systems and transmission capabilities, while the time available for data transfer is further restricted by the satellite's visibility window. Performing computations directly onboard eliminates the need to download raw data or upload processed results, thereby significantly enhancing the real-time responsiveness of observational operations. Second, implementing MLLMs in this context allows us to evaluate their operational feasibility and computational performance in space environments. This testbed serves as a critical stepping stone toward deploying more advanced scientific large language models on future satellite missions.

In this paper, we present the design and implementation of an in-orbit GRB identification framework based on machine learning, developed for the CXPD CubeSat. The remainder of the paper is structured as follows. Section~\ref{sect:Obs} describes the construction of the dataset, including background modeling and GRB spectral simulations. Section~\ref{sec:model} details the model architecture, training methodology, and deployment considerations for onboard execution. Validation results and in-orbit simulation tests are presented in Section~\ref{sect:result}. Finally, Section~\ref{sect:discussion} discusses the implications of our findings and summarizes the main conclusions.

\section{Dataset Preparation}
\label{sect:Obs}
\begin{figure}
   \centering
   \includegraphics[width=8cm, angle=0]{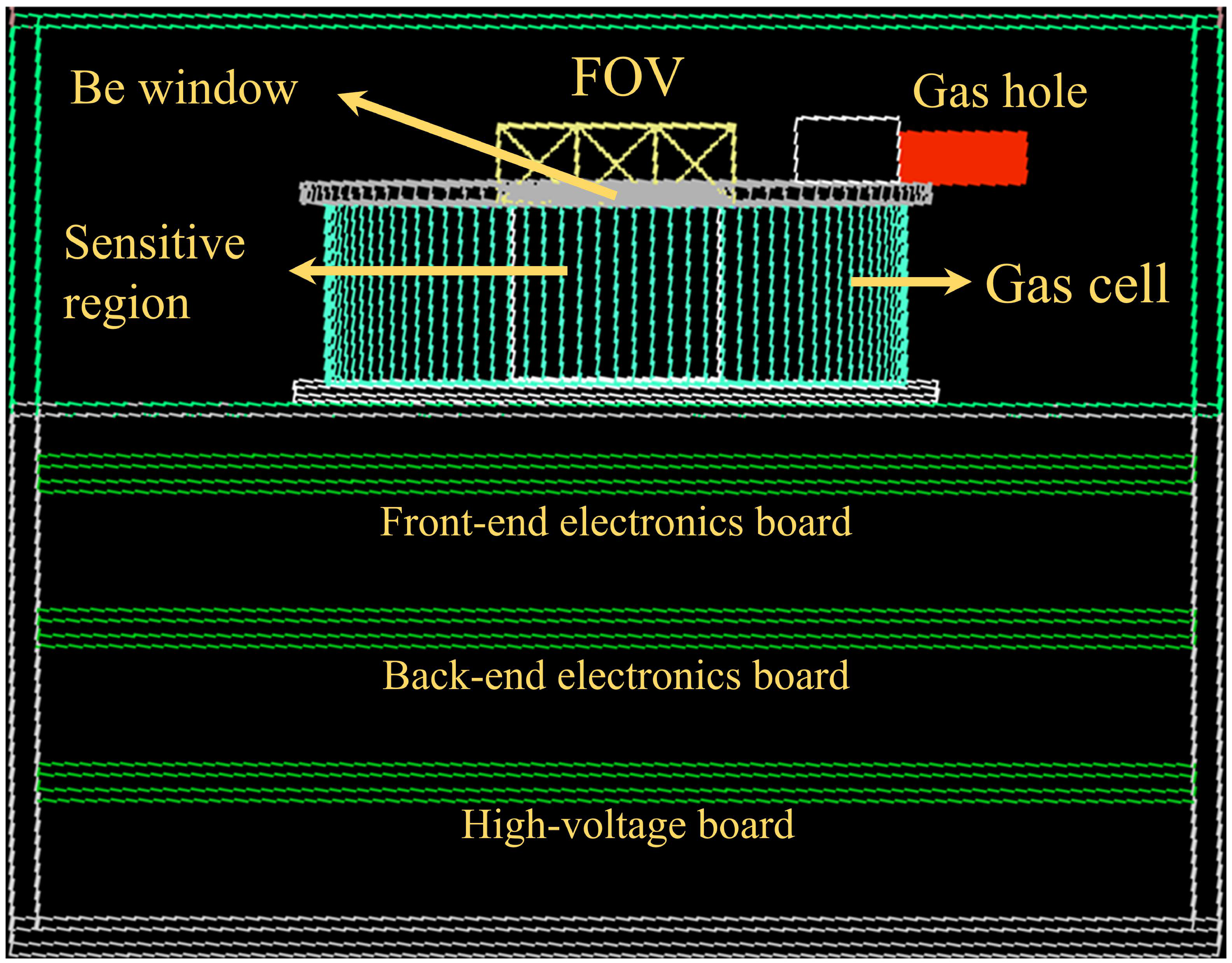}
   \caption{Mass model of CXPD-03 cubesat }
   \label{Fig1}
   \end{figure}

We employ the Geant4-based simulation framework star-XP, developed in-house, to generate both background and gamma-ray burst training datasets. The simulation utilizes a cubic satellite mass model as input, as illustrated in Figure~\ref{Fig1}.

\subsection{In-orbit Background}
CXPD-03/04 are expected to be launched into low Earth orbit at an altitude of approximately 500 km. Comprehensive studies of the background components affecting photoelectric polarimeters in low Earth orbit have been performed by \cite{xie2021study} and \cite{huang2021modeling}. Furthermore, \cite{Feng_2024} specifically examined the dominant in-orbit background sources relevant to wide FOV photoelectric polarimeters. Their analysis identified the primary background contributions as the cosmic X-ray background (CXB), charged particle background, and bright astrophysical X-ray sources—particularly those concentrated near the Galactic plane. The charged particle background can be substantially suppressed by leveraging both the energy information and track morphology of individual events, yielding a rejection efficiency of up to 90\%.

Additionally, the detector is programmed to reduce its high voltage when the Sun enters the field of view or when the CubeSat passes through regions of enhanced radiation, such as the South Atlantic Anomaly and charged particle belts near the polar regions. This precaution is designed to protect the detector from potential damage caused by intense solar X-ray fluxes and elevated charged particle radiation in these zones.

Building on this understanding of the background environment, the energy spectra of various background components are incorporated as inputs in the simulations. Then a sky survey simulation is conducted using the simulator, in which a total of 12,288 incident directions are modeled to represent the full spatial detection coverage (see \citealt{Feng_2024} for simulation details). The resulting response spectra from these directions are collected as background input samples for subsequent machine learning applications.

   \begin{figure}
   \centering
   \includegraphics[width=7 cm, angle=0]{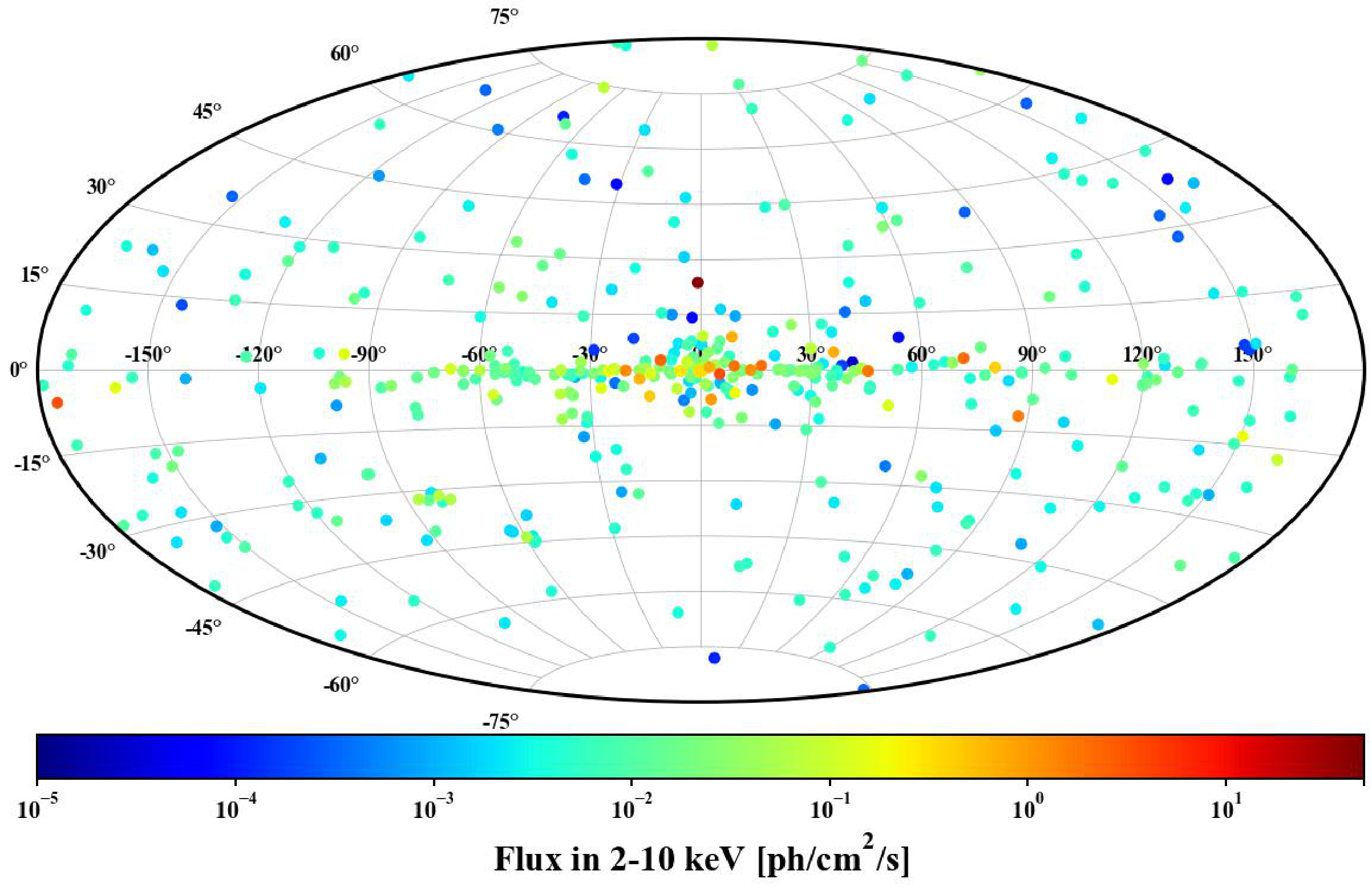}
   \includegraphics[width=7.5 cm, angle=0]{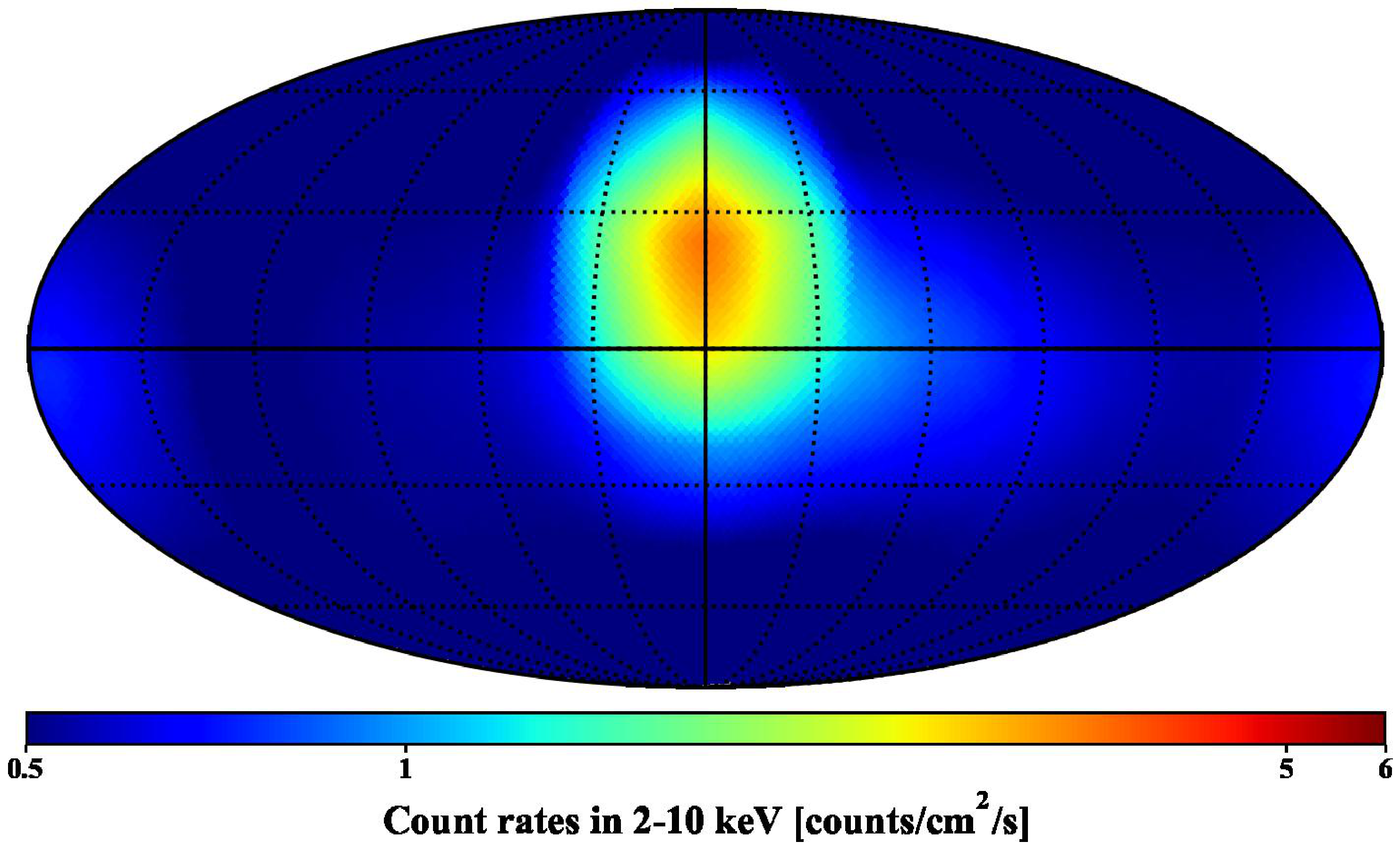}
   \caption{The left panel shows the source catalog constructed from MAXI observational data. The right panel presents the simulated sky survey background map, incorporating contributions from all MAXI/GSC sources, CXB, and charged particles. The map is plotted in Galactic coordinates (with longitude increasing from left to right) and consists of 12,288 pixels, each representing a distinct incident direction.}
   \label{Fig2}
   \end{figure}

\subsection{GRB}
\begin{figure}
   \centering
   \includegraphics[width=12cm, angle=0]{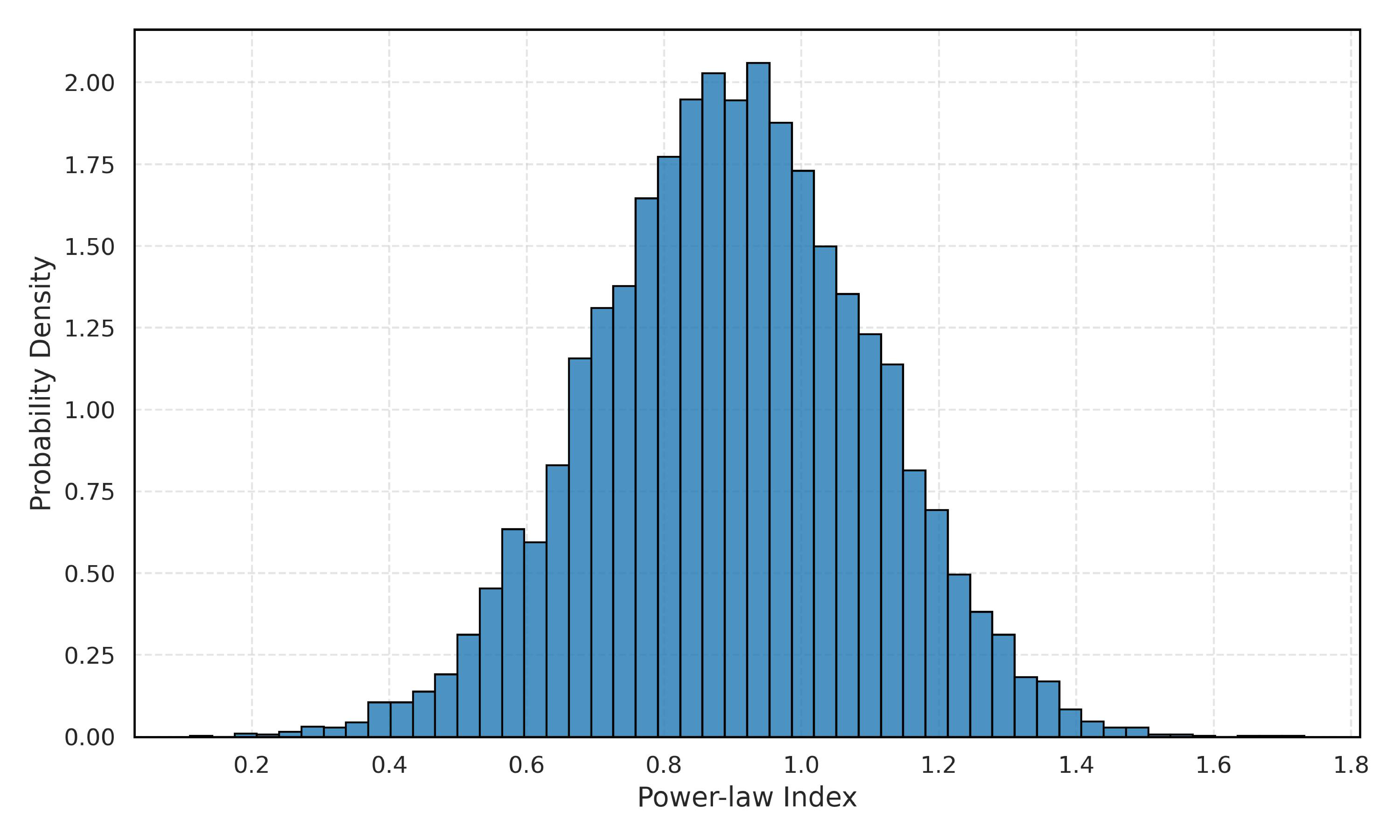}
   \caption{Distribution of the GRB photon index in the 2–10 keV energy range, based on statistical analysis of observations from the Swift satellite. This distribution is used for sampling the spectral indices in the GRB detection simulations.}
   \label{fig:Fig3}
   \end{figure}

For the GRB spectral input in the simulator, we focus on the energy range of 2–10 keV, which is particularly relevant for the study of  GRB prompt emission in the soft X-ray band. Within this band, the GRB spectrum is modeled as a simple power-law distribution. The photon indices are sampled according to the distribution obtained from statistical analyses of Swift satellite observations \citep{lien2016third}, as shown in Figure \ref{fig:Fig3},  which is characteristic of prompt-emission-dominated spectra.

During in-orbit operations, GRB identification is scheduled to occur every 300 seconds. This time window is determined by the operational design of the instrument and is intended to represent the extended duration of GRB prompt emission in the soft X-ray band, which is known to be significantly longer than that observed in the gamma-ray band and can reach several hundred seconds, as indicated by observations from the Einstein Probe (EP). Based on the GRB flux distribution reported in statistical studies and using the previously described spectral samples, we generate the cumulative GRB spectrum for each 300-second interval as input to the simulation. These simulated spectra are then used to model the detector's response to prompt-emission-dominated GRB signals.

For the energy spectra used in the machine learning framework, only GRB events with total count rates exceeding the average background level are included. This selection criterion ensures that only detectable events are considered, with an emphasis on bright GRBs for which the prompt emission is expected to dominate the observed flux during the initial few hundred seconds in the 2–10 keV band. We note that for certain GRBs, particularly those with rapidly fading prompt emission or a strong early afterglow component, the accumulated spectrum over 300 seconds may include a non-negligible contribution from afterglow emission. However, the detailed modeling of afterglow spectral and temporal evolution is beyond the scope of this work and is therefore not explicitly included in the simulations.

\section{Methodology} \label{sec:model}

In this work, we aim to develop a robust on-orbit system for GRB triggering and identification using a MLLM. While high-precision spectral analysis will continue to rely on established ground-based methods, we incorporate photon index regression here not for accuracy, but as a probe of the MLLM’s capacity to extract meaningful physical information from onboard data. This auxiliary task helps assess the model’s potential for in-orbit scientific reasoning and informs the feasibility of deploying such architectures on resource-constrained satellite platforms. Our approach thus focuses on exploring the capabilities and practical implementation of large models in space, rather than replacing conventional spectral fitting techniques.

\subsection{miniCPM}
The miniCPM-V series \citep{miniCPM2.6git, miniCPM2.6paper} comprises edge-side MLLMs that can be deployed on mobile devices such as smartphones, offering image and video processing capabilities comparable to those of GPT-4V \citep{openai2024gpt4technicalreport}. Notably, miniCPM-V 2.6, an open-source model with a total of 8 billion parameters, surpasses GPT-4V in image understanding tasks. This version has achieved real-time video understanding on an iPad for the first time, thanks to its superior token density \citep{miniCPM2.6git}.

Deploying models on satellites presents significant challenges, primarily due to limited computational resources and restricted data transfer rates between the satellite and Earth. In this context, the miniCPM series stands out, particularly because it excels at model compression while maintaining robust image processing capabilities. Consequently, we have selected miniCPM-V 2.6 for our application, aiming to classify GRBs from background noise and regress the power-law index aboard the CXPD. For simplicity, we refer to miniCPM-V 2.6 as \emph{miniCPM} throughout this paper.

\subsection{Training set design}
The simulation described in Section \ref{sect:Obs} is used to construct the training set. The primary objective of our spaceborne GRB model is to evaluate the feasibility of edge computing on satellites and to demonstrate a comprehensive model for both GRB identification and power-law index regression \citep{Dband1993GRB}.

The CXPD simulation dataset consists of two main components: 69,993 GRB signals with varying spectral indices, and 12,288 background signals. Each signal is transformed into an energy spectrum diagram divided into 20 bins (for details on the distribution of GRB and background signals, including sample counts and spectral indices, refer to Figure~\ref{fig:Fig3}).

The training set is subsequently partitioned into a training subset and a validation subset. For GRB data, the split ratio is 4:1, while for background signals, it is 6:1, as background signals do not require spectral regression. This division ensures that the training subset contains sufficient data to accurately represent the underlying distributions. Specifically, the training subset comprises 55,993 GRBs and 10,531 background objects, totaling 66,524 objects. The validation subset includes 14,000 GRBs and 1,756 background objects, summing up to 15,756 objects.

\begin{figure}
    \centering
    \includegraphics[width=0.45\linewidth]{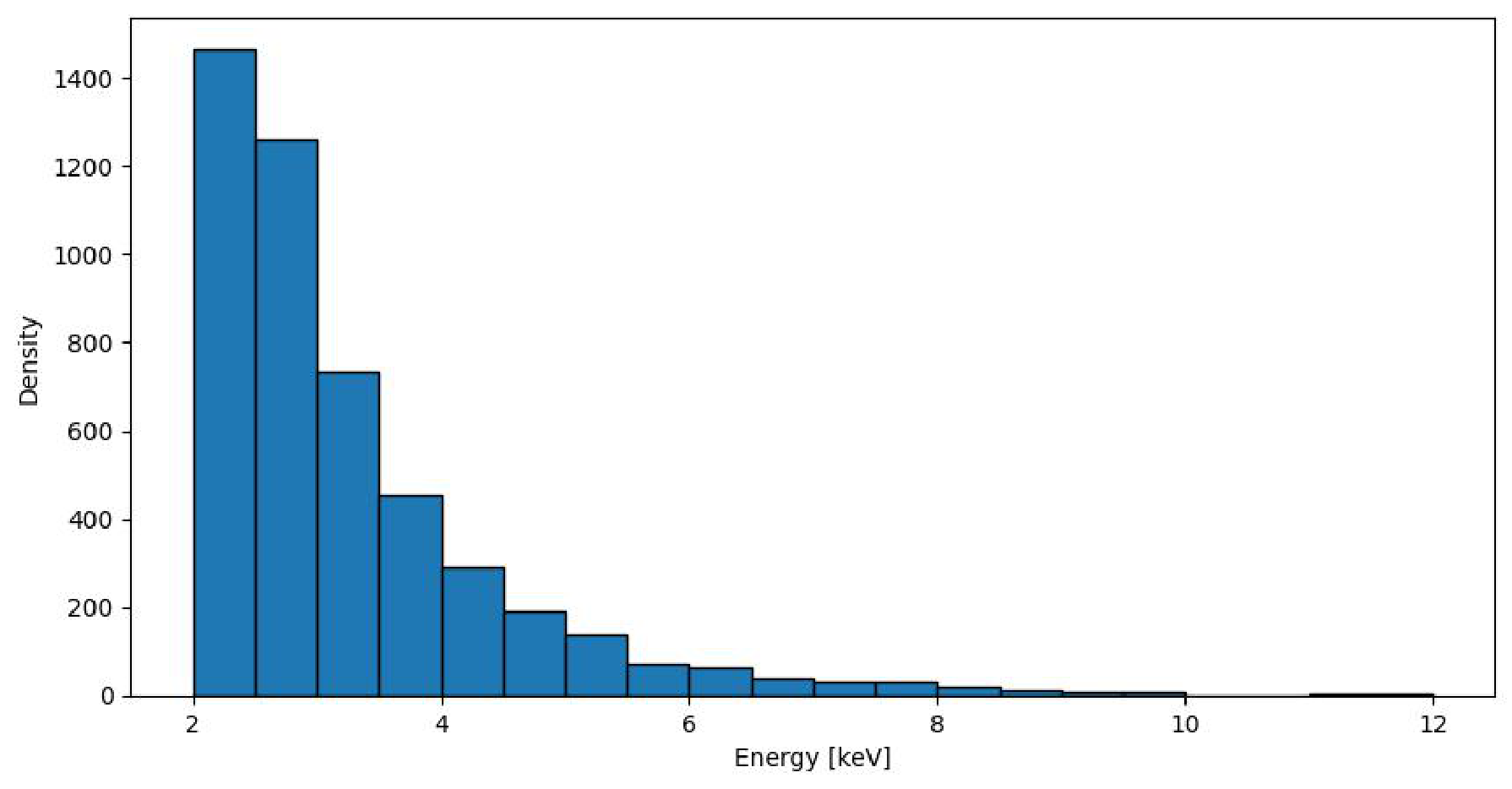}
    \includegraphics[width=0.45\linewidth]{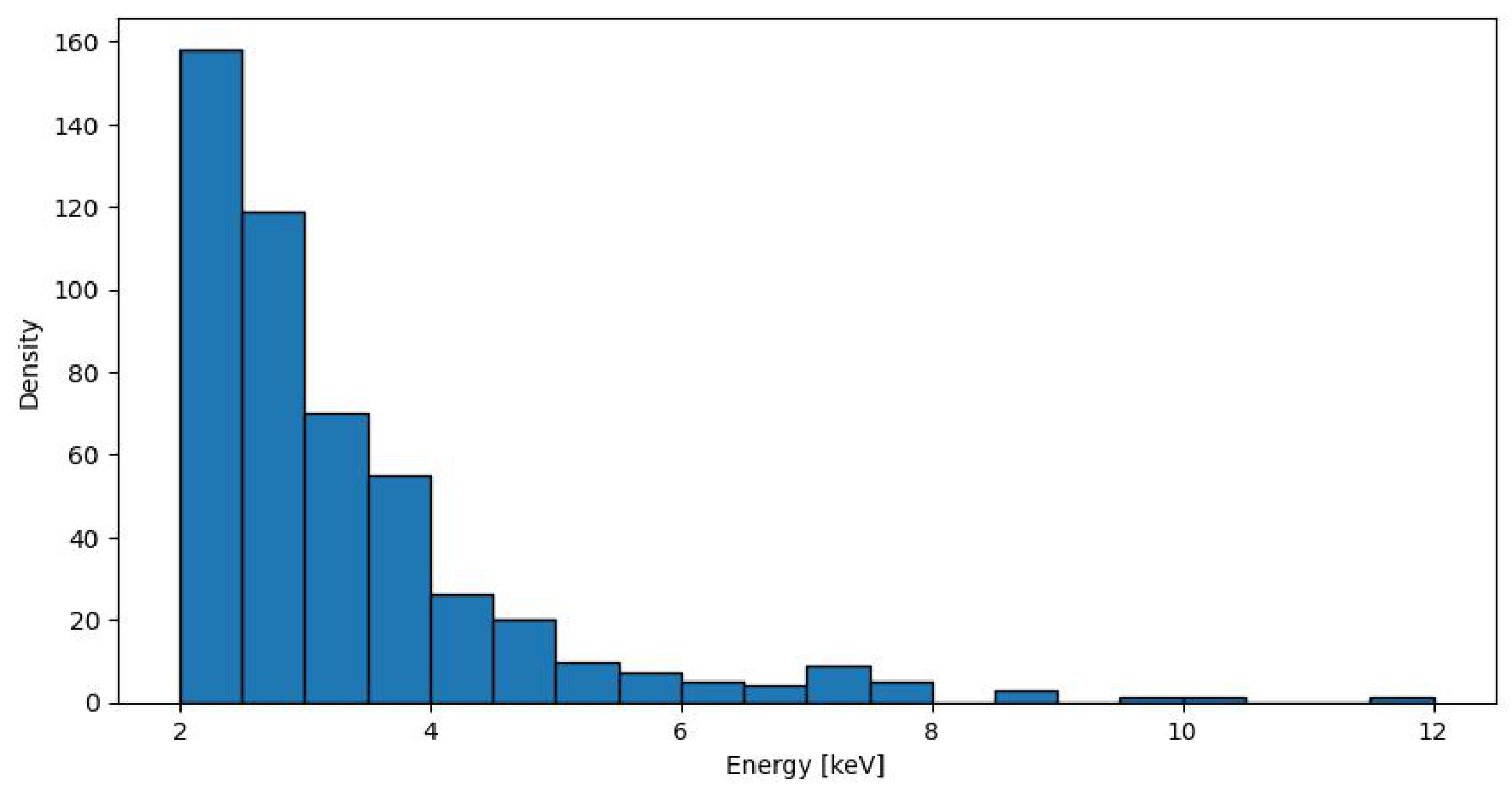}
    \caption{The energy-spectrum diagram for a GRB with power-law index equal to 1.224 (left panel) and a background (right panel) in the training set.}
    \label{fig:sample_diagram}
\end{figure}

\citet{zhu2025grbclassfy} applied energy-spectrum diagrams to do their GRB classification with machine learning methods. Considering the energy-spectrum diagram is easy to obtain and needs little calculation cost, we also choose it as the input of our in-orbit model.  Figure~\ref{fig:sample_diagram} illustrates an energy spectrum diagram utilized in our model training process.

The design of the input image alone is insufficient to train a well-performing model \citep{wei2023chainofthoughtpromptingelicitsreasoning}. Therefore, we also designed a structured prompt to guide the model’s reasoning process (see Prompt~\ref{prompt:input-prompt}). In this prompt, the content within curly braces \{\} represents parameters dynamically filled in based on prior information.

For instance, \emph{spectral\_energy\_diagram} refers to an array containing the actual values of the 20 bins from the energy-spectrum diagram histogram, with each value rounded to six decimal places. The \emph{power-law index} denotes the power-law index of the corresponding object in the training set, which is set to 0 for background objects, as they do not exhibit significant spectral features.

\begin{lstlisting}
#query
    This is an X-ray energy spectrum diagram. Please determine if it is a gamma ray burst (GRB) or a background. 
    - If it is a GRB, reply with 'GRB'. 
    - If it is a background, reply with 'Background'. 
    Additionally, provide the following information:
    - Spectrum index (if it is a GRB): 'specIndex: **the spectrum index**'
    - If it is not a GRB, set the spectrum index to 0.
    All numbers should be separated by spaces.
    The value of the spectrum diagram with 20 bins are 
    '''{spectral_energy_diagram}''' 
    Please think step by step and provide the result.
    
#response
classification: GRB (or Background), specIndex: '''the power-law index''' 
\end{lstlisting}
\captionof{lstlisting}{The query-response is the standard format to train a MLLM. The spectral\_energy\_diagram is the bin value of the spectral energy diagram. All numerical data in the prompt is separated with a space in each decimal.\\}
\label{prompt:input-prompt}

In our dataset, all numerical values are formatted with spaces separating each digit and decimal point. For example, the number 9.11 becomes "9\quad.\quad1\quad1". This formatting strategy enhances the model’s ability to distinguish individual numerical components, ensuring that numbers are interpreted as such rather than being misclassified as dates, labels, or other symbolic representations during tokenization.

This issue has been observed in prior studies involving LLMs, where models such as Llama 3.1 8B have incorrectly interpreted numerical comparisons—such as determining whether 9.11 is greater than 9.8—due to contextual associations (e.g., linking "9.11" to terrorist attacks or biblical chapters, \citealt{choi2024automatic}\footnote{\url{https://transluce.org/observability-interface}}). Research by \citet{ghandeharioun2024911} demonstrates that spacing the digits significantly improves the model's numerical reasoning capabilities by reducing ambiguity and allowing clearer separation of components.

Although our work involves spectral data rather than textual or symbolic content, we cannot rule out potential misinterpretations during the tokenization process. Therefore, we adopt this spaced-digit format to mitigate possible parsing errors and ensure accurate numerical interpretation by the model.

\subsection{Training} \label{sec:auto}

The training process is conducted using Swift\footnote{\url{https://github.com/modelscope/ms-swift}} to perform supervised fine-tuning (SFT) with the aid of the \textbf{LoRA} method \citep{hu2021lora}. Training was carried out on a server equipped with two NVIDIA RTX 4090 GPUs, providing a total of $2 \times 24$GB of Video Random Access Memory (VRAM).

LoRA is an efficient fine-tuning technique that enables model adaptation at a significantly reduced computational cost, without modifying the base model parameters. It introduces a low-rank matrix into the cross-attention layer weights, which is then decomposed into two smaller matrices to minimize storage requirements. We selected LoRA as our training strategy for two key reasons: (1) The available VRAM can support LoRA-based training but not full-parameter fine-tuning; and (2) LoRA modules can be combined or switched to generate different outputs, offering flexibility in model behavior.

The resulting training and validation datasets are formatted as JavaScript Object Notation Lines (.jsonl) files, where each line contains a structured triplet of query, response, and images—a standard input format for MLLM training using Swift.

To meet the edge-computing constraints of satellite deployment, the trained model was quantized to 4 bits with the help of bitsandbytes\footnote{\url{https://github.com/bitsandbytes-foundation/bitsandbytes}}. Model size reduction through quantization enhances both deployment efficiency and operational stability in resource-limited environments. However, quantization inherently involves a trade-off between inference accuracy and memory footprint. Since training is performed on a ground-based GPU server, we are able to prioritize accuracy by training over two epochs, thereby leveraging the full representational capacity of the dataset. See Appendix \ref{app:training details} for the training command via Swift.

Figure~\ref{fig:training} presents the training loss and token accuracy throughout the training process. Notably, the model converges rapidly within the early stages of the first epoch.

\begin{figure}
    \centering
    \includegraphics[width=0.45\linewidth]{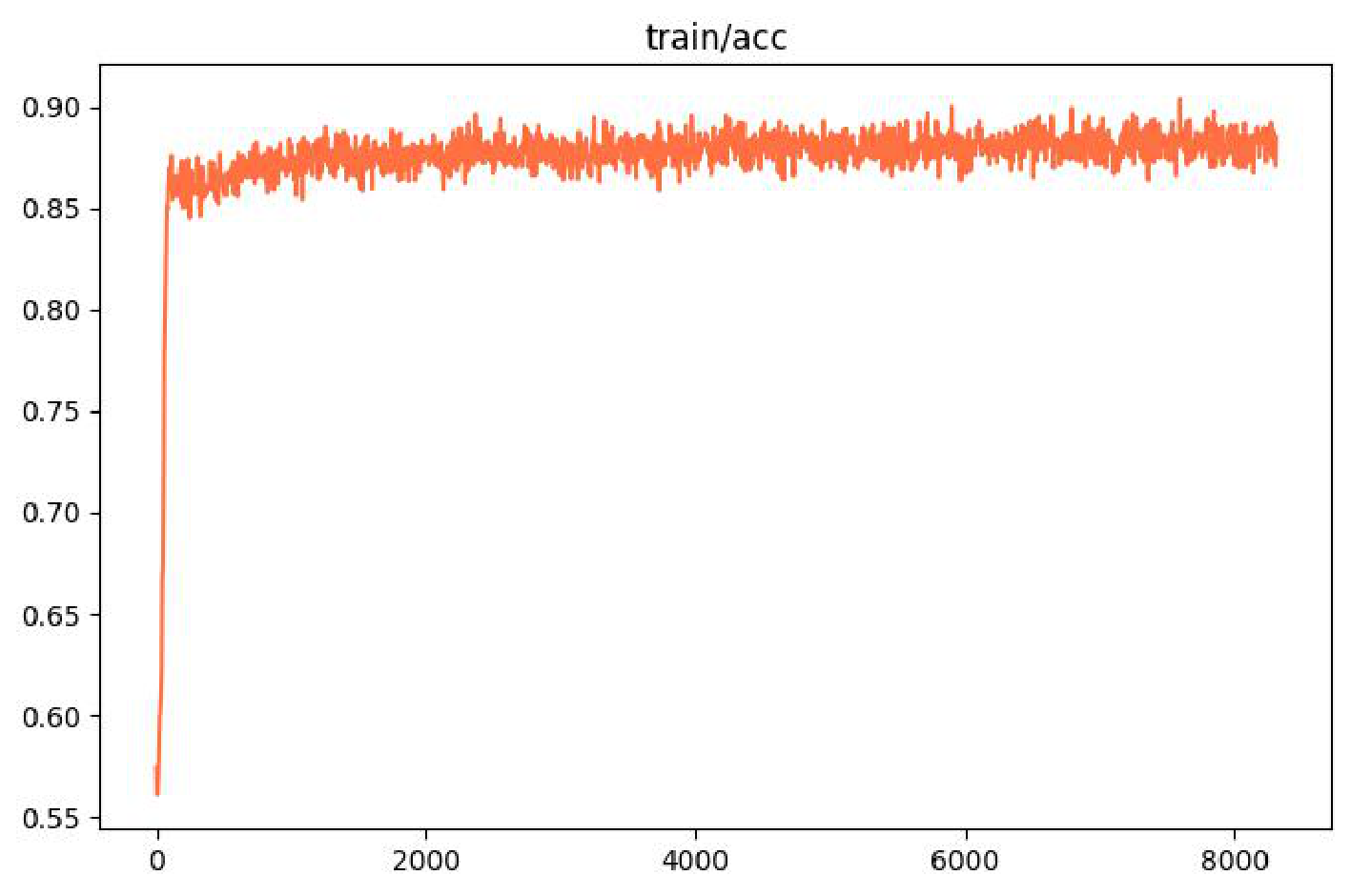}
    \includegraphics[width=0.45\linewidth]{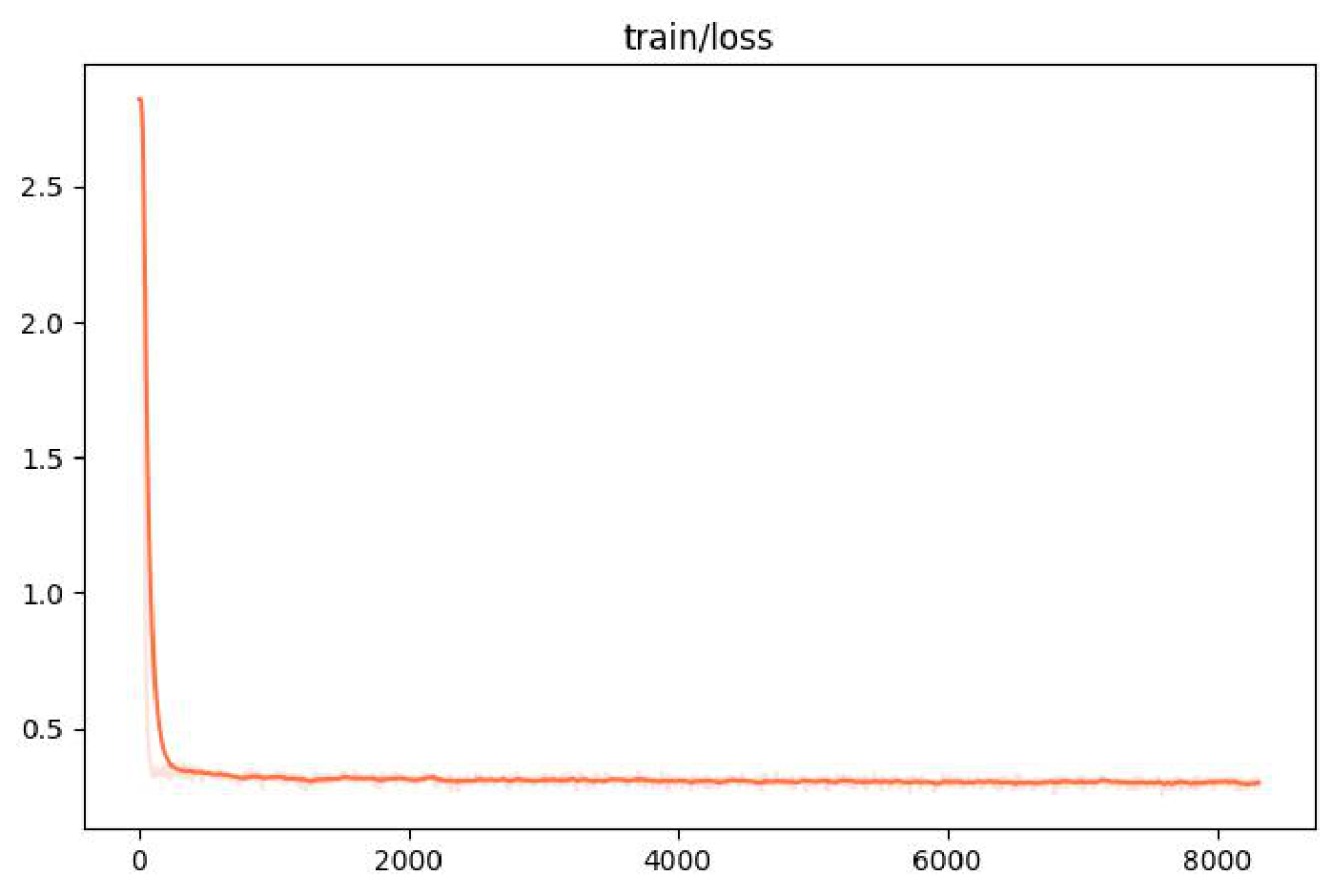}
    \caption{The left panel is the training accuracy, and the right panel is the training loss.}
    \label{fig:training}
\end{figure}

\section{Result}
\label{sect:result}

\subsection{Model validation}
\begin{figure}
    \centering
    \includegraphics[width=0.6\linewidth]{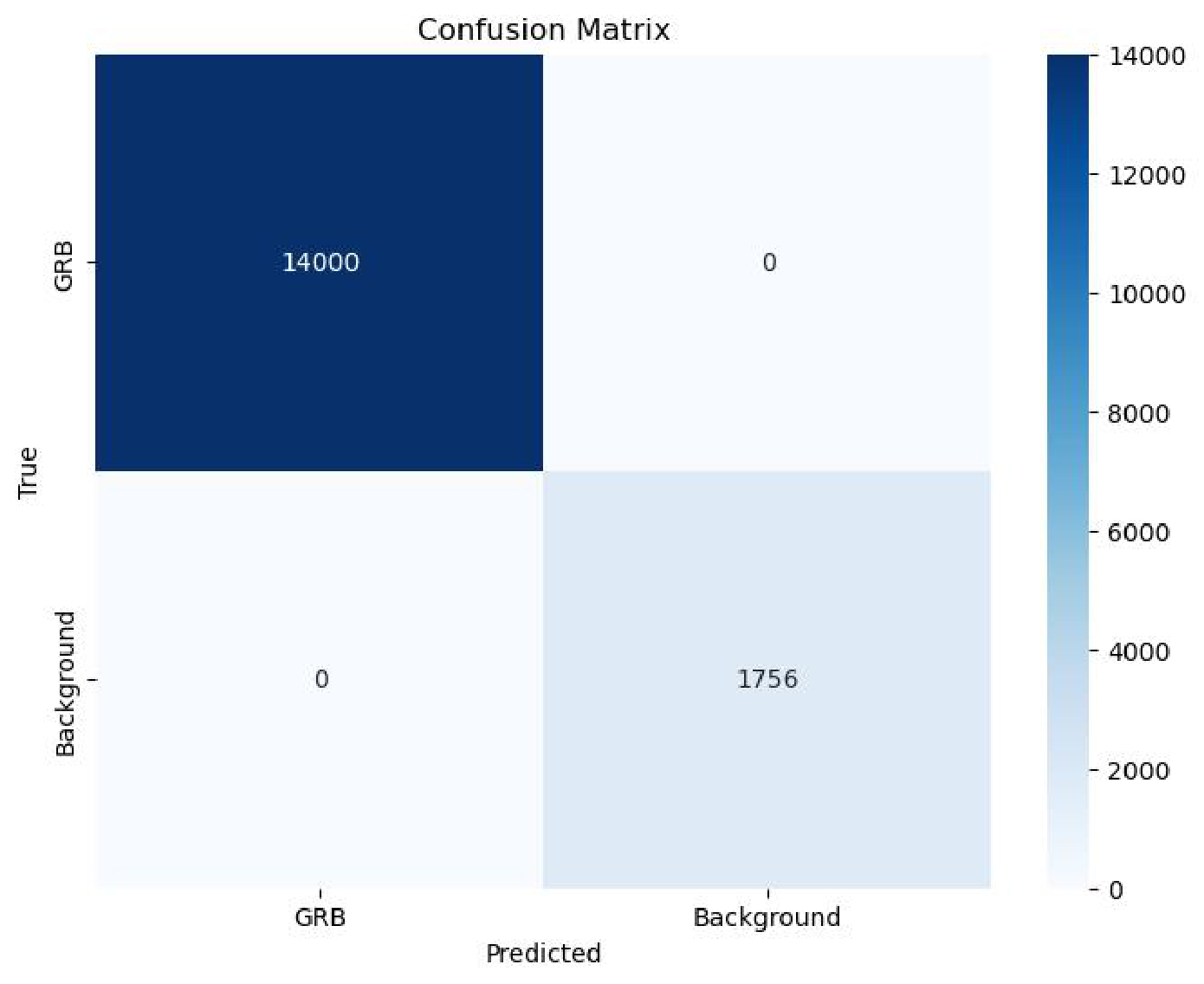}
    \caption{The confusion matrix for the classification.}
    \label{fig:confmatrix}
\end{figure}

We evaluated the model using the validation set derived from the simulation data. The evaluation aims to assess two primary tasks: (1) Classification of GRB signals against background signals; (2) power-law index regression for GRB signals. The classification accuracy achieved is 1 (or 100\%), indicating perfect discrimination between GRB and background signals. This high accuracy can be attributed to the significant differences between GRB signals and background noise. The confusion matrix illustrating this classification performance is shown in Figure~\ref{fig:confmatrix}.

For the power-law index regression task, only GRB data are used in the validation process. This is because the background samples are assigned a fixed power-law index value of 0 during training. Furthermore, the validation set contains no GRB samples with a power-law index of 0, nor any background samples with non-zero spectral indices. This design choice demonstrates that, even in the context of MLLMs, the regression task exhibits behavior consistent with the classification task—namely, a clear separation between GRB and background signals.

\begin{figure}
    \centering
    \includegraphics[width=0.6\linewidth]{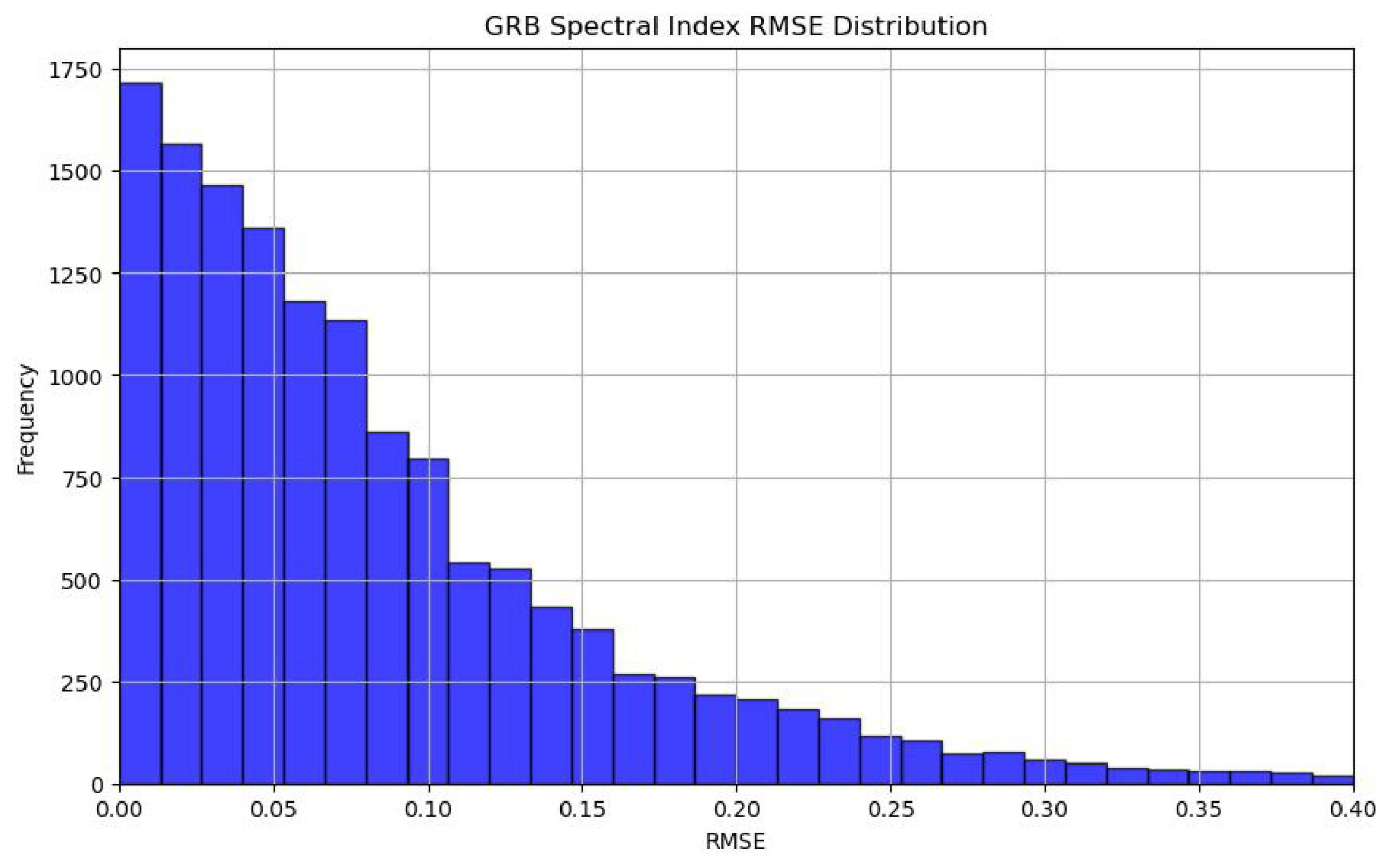}
    \caption{The RMSE distribution of the regression task on the GRB power-law index regression.}
    \label{fig:rmsedis}
\end{figure}

The average root mean square error (RMSE) for the power-law index regression is 0.118, with most predicted values falling within an RMSE range of 0 to 0.4, as shown in Figure~\ref{fig:rmsedis}. To provide further insight into the regression performance, we also present a one-to-one prediction vs. ground truth plot, along with the error distribution histogram, in Figure~\ref{fig:training}. The mean prediction error is less than 0.01, and the standard deviation of the error is approximately 0.118, indicating both high accuracy and consistency in the model's regression capability.

\begin{figure}
    \centering
    \includegraphics[width=0.45\linewidth]{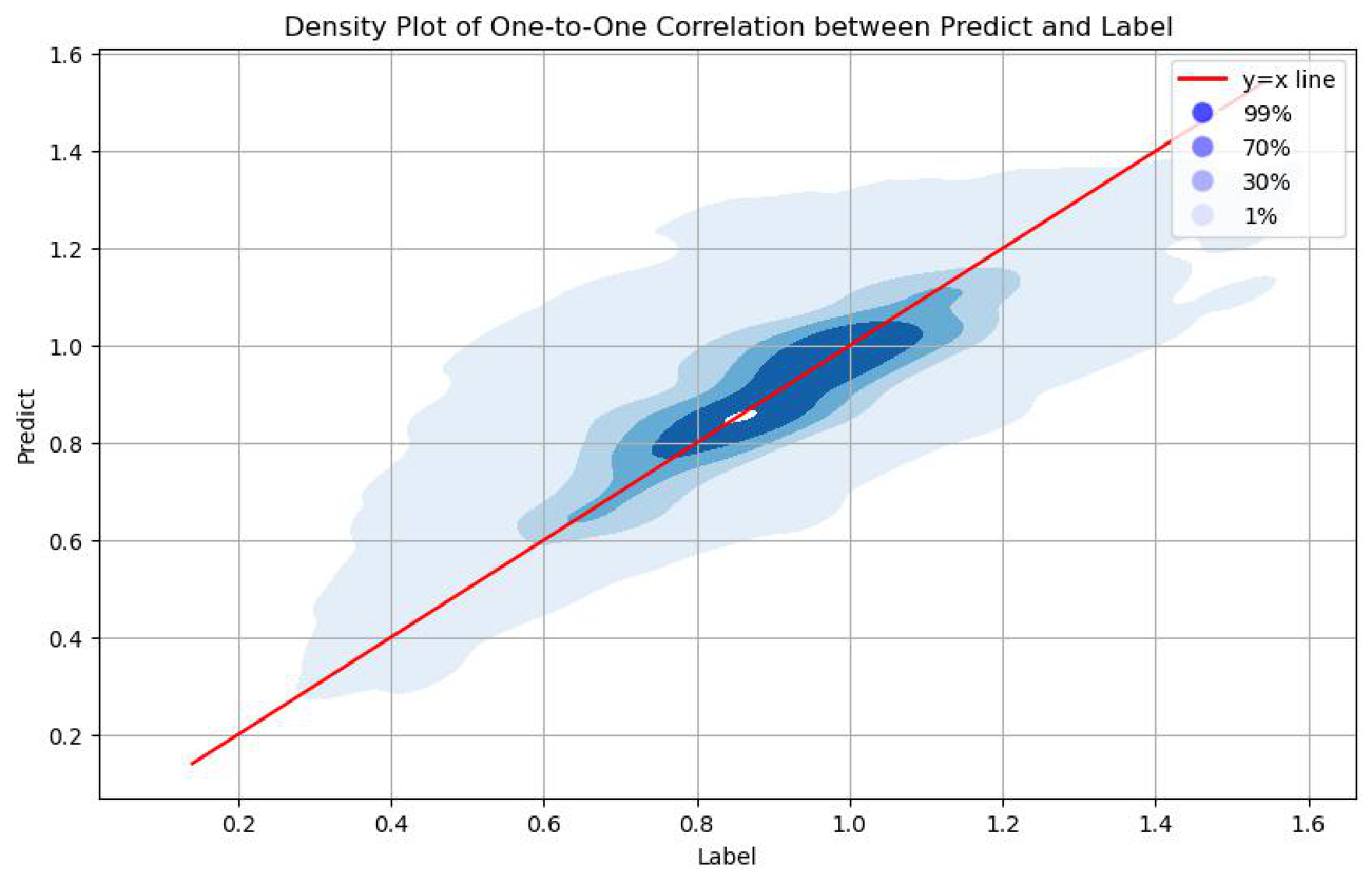}
    \includegraphics[width=0.45\linewidth]{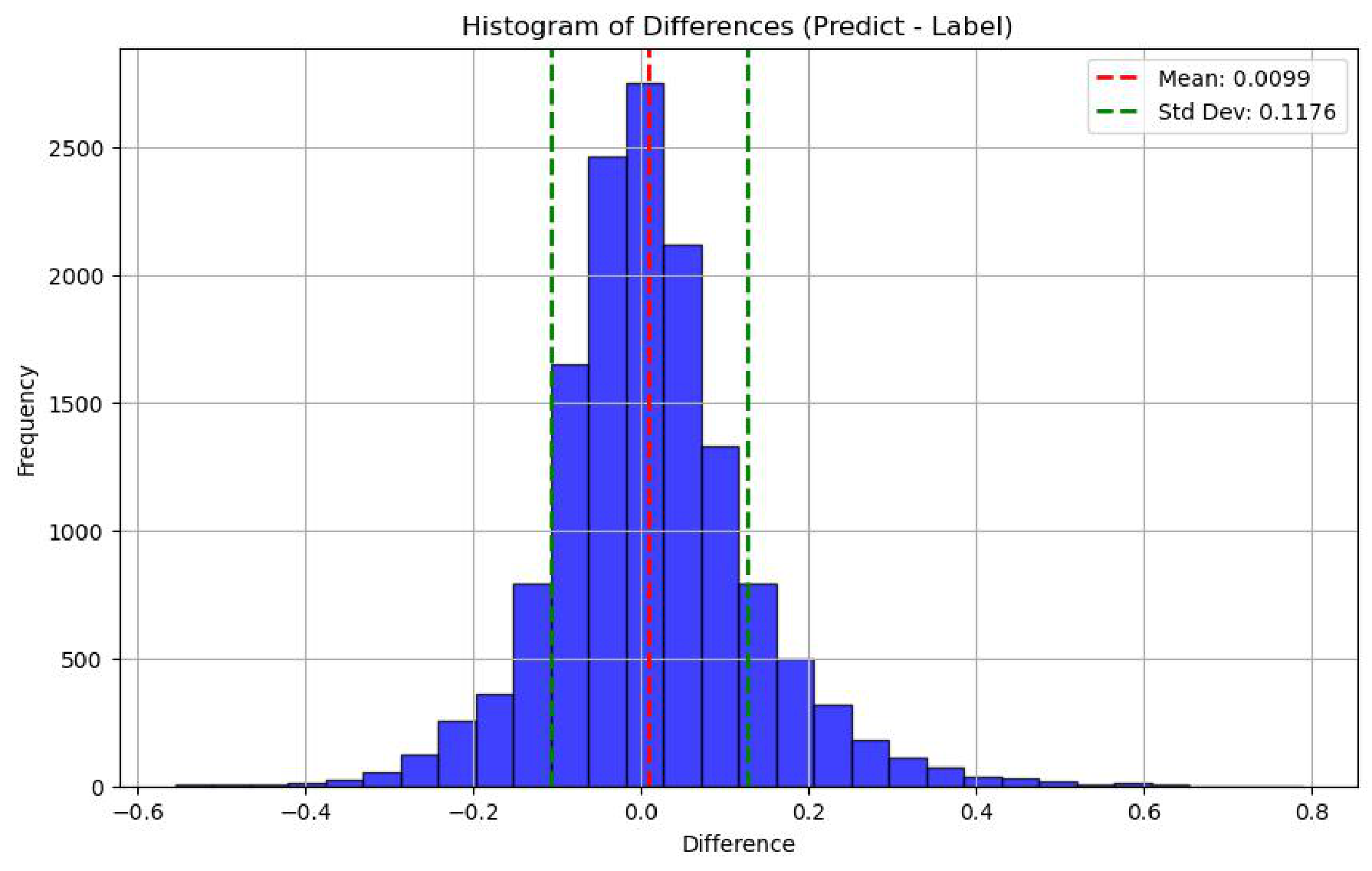}
    \caption{The left panel is the one-to-one correlation diagram, and the right panel is the training error distribution.}
    \label{fig:training}
\end{figure}

For comparison, we also implemented a lightweight multi-task MLP (details in Appendix~\ref{app:appcnn}) to jointly classify GRBs versus background and regress the spectral index. The model achieves 98.16\% overall accuracy and 99.21\% GRB recall, but yields a relatively high regression RMSE of 0.1815, worse than the MLLM. Notably, it produces physically implausible positive spectral indices for 2,455 out of 2,458 background events, suggesting limited capacity in capturing underlying physical constraints. This highlights that conventional small-scale architectures, while computationally efficient, may lack the representational power needed for robust multi-task astrophysical inference.

\subsection{On Satellite pipeline}
We conducted ground-based simulations of GRB observations and onboard data processing to evaluate our model's performance. These simulations replicated the observational conditions of a CubeSat in orbit, focusing on GRBs within a continuous X-ray spectrum range of 2-10 keV.

During the observation phase, the CubeSat collects various types of monitoring data, including timestamps, detector temperature, pressure, voltage, and current. Additionally, it records scientific data such as pulse signal intensity on the Gas Microchannel Plate (GMCP) lower surface and chip-triggered pixel values. All of these data are stored in a highly compressed binary format within the onboard data storage unit to optimize memory usage.

Upon completion of the observation period, the CubeSat transitions to a data transmission state. Raw observation data are transmitted to an in-orbit computing platform via an inter-satellite communication system. On this computing platform, we have implemented a C++-based data parsing and processing software framework designed to interpret the raw CubeSat data. This framework converts engineering data into commonly used analysis formats, such as CERN ROOT and PYTHON numpy files.

For track data, the parsing framework employs erosion and clustering algorithms to remove noise from images and separate clusters formed by different tracks. For GMCP lower surface data, the framework translates pulse intensity into photon energy information based on pre-calibrated ground calibration results (specific energies: 2.98 keV, 4.51 keV, 5.4 keV, 5.89 keV, 6.40 keV, and 8.04 keV). It also corrects these values according to real-time chamber temperature monitoring data collected during the observation.

Finally, the parsing framework generates energy spectrum images over a specified interval—defaulting to 300 seconds—as input for subsequent machine learning models. These processed datasets facilitate accurate GRB detection and power-law index regression, thereby enhancing the CubeSat's ability to perform edge computations in space environments.

\section{Conclusion}
\label{sect:discussion}
In this study, we propose and validate an innovative method for in-orbit GRB identification using the CXPD CubeSat and an MLLM. Leveraging the miniCPM-V 2.6 model, we successfully classify GRB signals and regress their spectral indices from simulated energy spectrum diagrams within the 2–10 keV range. To optimize the model for onboard deployment under constrained computational resources, we fine-tuned it using the LoRA technique and quantized it to 4-bit precision.

Our experimental results demonstrate outstanding performance: a classification accuracy of 100\% and an RMSE of 0.118 for power-law index regression. These outcomes indicate that our model can effectively distinguish GRB events from complex and variable background conditions encountered in wide-field X-ray polarimetry. We note that the background class used in this study primarily consists of instrumental and diffuse background components. Some non-GRB astrophysical transients, if sufficiently bright, may still be of interest for polarization observations and are therefore not necessarily required to be excluded at the triggering stage.

To further verify the operational feasibility of our approach, we implemented a comprehensive simulated data acquisition and processing pipeline. This pipeline includes detector response simulation, engineering data parsing, and spectral image generation, thereby mimicking the complete onboard data flow. Our simulations confirm the practicality of deploying deep learning-based GRB identification models in resource-limited CubeSat environments.

This work underscores the potential of utilizing compact, edge-deployable MLLMs to enhance transient detection capabilities for space missions. Future research will focus on expanding the training dataset with more realistic orbital scenarios, improving regression precision for weak GRBs, and validating the system using in-flight data from upcoming CXPD missions.

The CXPD CubeSat was launched as part of the three-bodied computing constellations—a network of satellites designed for space-based cooperative computing—from the Jiuquan Satellite Launch Center on May 14, 2025, aboard a Long March 2D rocket. Subsequent tests will evaluate the capability of running LLM-based applications in space.

\begin{acknowledgements}
The research presented in this paper was generously funded by the National Programs on Key Research and Development Project, with specific contributions from grant numbers 2019YFA0405504 and 2019YFA0405000. This work was funded by the National Natural Science Foundation of China (NSFC) under NSFC-11988101, 11973054, 11933004, 11080922 and 12221005. We acknowledge the foundation from Zhejiang Province Key Research and Development Plan(No.2024SSYS0006) and lnnovation Project of Guangxi Graduate Education (No.YCBZ2025045).

We also received backing from the Strategic Priority Program of the Chinese Academy of Sciences, granted under XDB41000000. Special acknowledgment goes to the China Manned Space Project for their science research grant, denoted by NO.CMS-CSST-2021-B07. We also acknowledge the Science and Education Integration Funding of University of Chinese Academy of Sciences.

JFL extends gratitude for support received from the New Cornerstone Science Foundation, particularly via the NewCornerstone Investigator Program, and the honor of the XPLORER PRIZE.

\end{acknowledgements}

% \appendix                  %%appendicial material is supported

% \section{This shows the use of appendix}

\appendix
\section{Training Details}
\label{app:training details}

In this section, we detail the training procedure outlined in Listing~\ref{prompt:sft}. All experiments are conducted using Swift version 2.6, as input keyword conventions vary across versions and may impact compatibility. The training is configured to run on two GPUs without a validation set, a design choice motivated by computational efficiency—incorporating validation would at least double the total training time. Although validation is typically used to monitor generalization and enable early stopping, it is unnecessary in this case: the MLLM demonstrates strong inherent learning capabilities, and convergence is consistently achieved within two epochs, rendering early stopping redundant.

For the batch size, we adopt a standard empirical strategy: selecting the largest value that avoids out-of-memory (OOM) errors given the available GPU memory. The setting train\_dataset\_sample=-1 indicates that no subsampling is applied to the training set, while dataset\_test\_ratio=0 ensures no train-validation split is performed during training. This is because a dedicated validation set was already established prior to training. All other hyperparameters retain their default values as specified in the official Swift repository (\url{https://github.com/modelscope/ms-swift}).

Inference is carried out on the satellite using Swift, as shown in Listing \ref{prompt:infer}. The process requires only two inputs: the paths to the trained model and the dataset. Notably, the onboard platform provides just one GPU, which is allocated to support the GRB detection model during inference.

\begin{lstlisting}
# Set environmental variables
export CUDA_VISIBLE_DEVICES=0,1
export NPROC_PER_NODE=2
export MODELSCOPE_CACHE= xxx
export OUTPUT_DIR= xxx

# run SFT command
swift sft \
  --model_type minicpm-v-v2_6-chat \
  --model_id_or_path OpenBMB/MiniCPM-V-2_6 \
  --sft_type lora \
  --dataset {Path to training jsonl file} \
  --train_dataset_sample -1 \
  --num_train_epochs 2 \
  --sft_type lora \
  --deepspeed default-zero2 \
  --quantization_bit 4 \
  --dataset_test_ratio 0 \
  --save_total_limit -1 \
  --save_strategy steps \
  --max_length 8192 \
  --per_device_train_batch_size 8
\end{lstlisting}
\captionof{lstlisting}{The full training command using swift.}
\label{prompt:sft}

\begin{lstlisting}
# Set environmental variables
export CUDA_VISIBLE_DEVICES=1
export NPROC_PER_NODE=1

# run SFT command
swift infer \
    --ckpt_dir {Path_to_checkpoint} \
    --load_dataset_config true \
    --val_dataset {Path_to_the_automatically_generated_dataset} \
    --show_dataset_sample -1
\end{lstlisting}
\captionof{lstlisting}{The code for the inference on the setallite.}
\label{prompt:infer}

\section{Comparison with MLP}
\label{app:appcnn}

We also trained a multi-task deep neural network based on a fully connected architecture (MLP) to simultaneously perform binary classification (distinguishing GRBs from background sources) and regression (predicting the spectral index). The model consists of three hidden dense layers with 128, 64, and 32 units, respectively, each using ReLU activation and 30\% dropout for regularization. These layers feed into two separate output heads: a softmax layer for classification and a linear layer for regression. The entire network is trained end-to-end using the Adam optimizer with a learning rate of $10^{-3}$.

\begin{figure}
    \centering
    \includegraphics[width=0.45\linewidth]{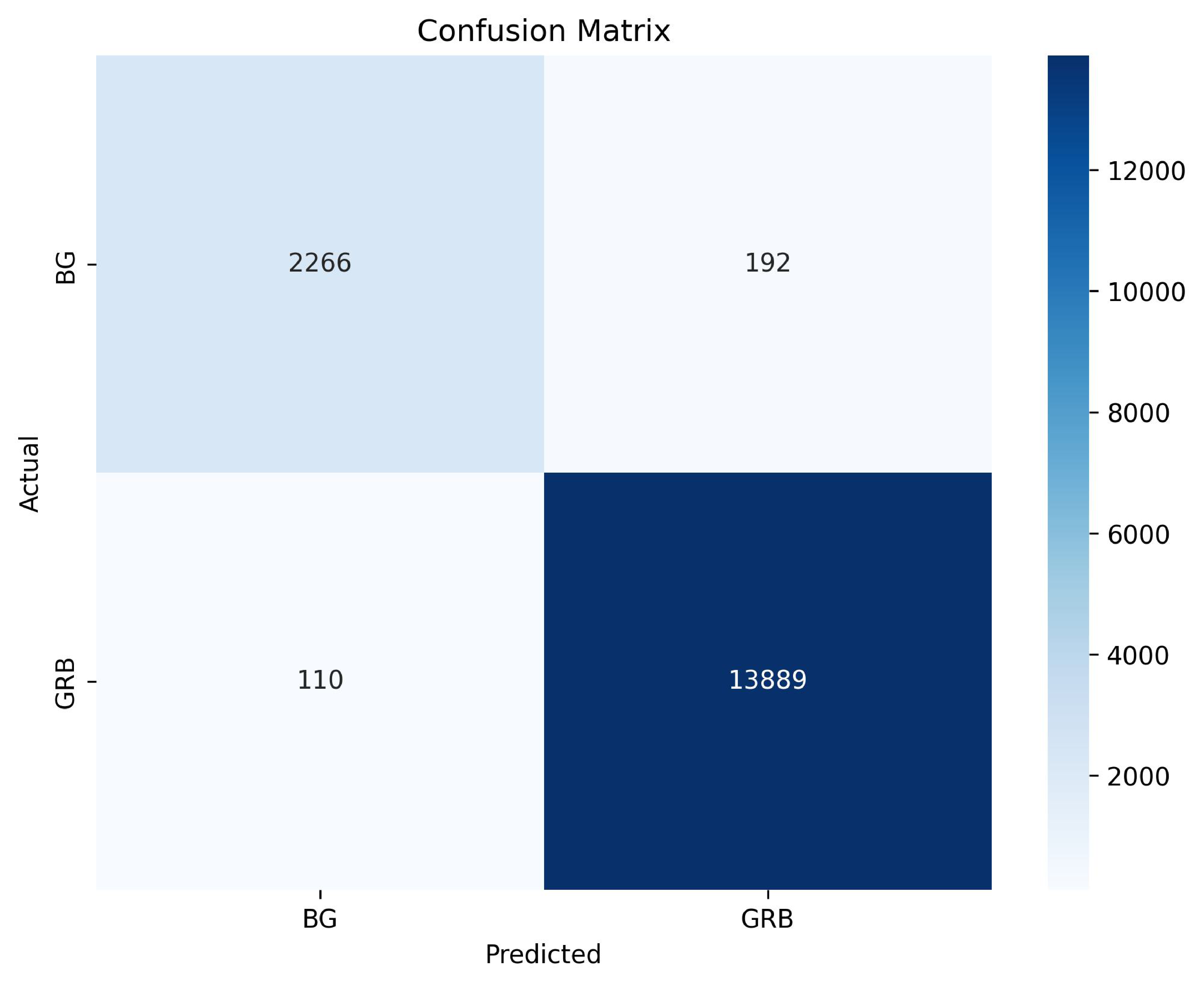}
    \includegraphics[width=0.45\linewidth]{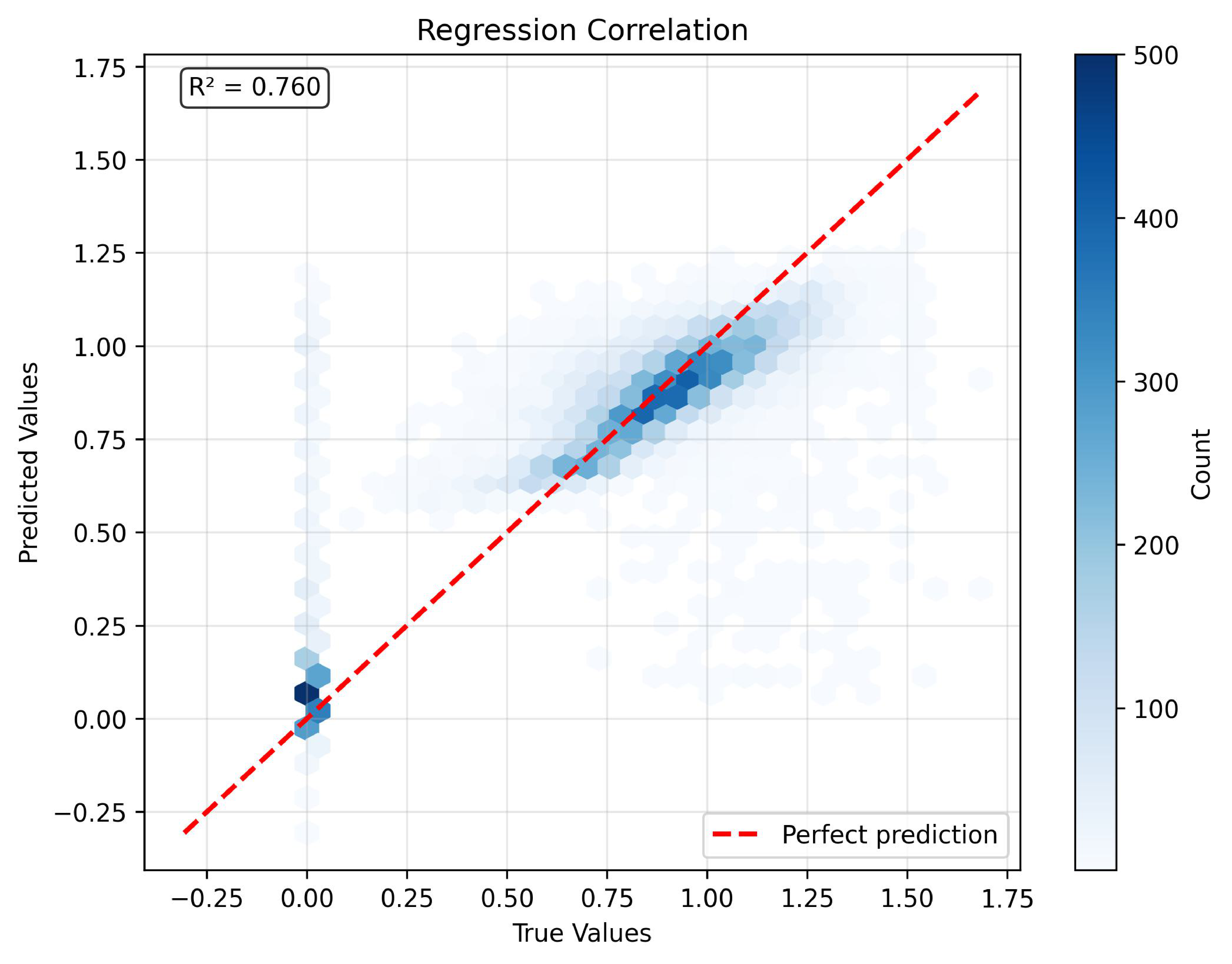}
    \includegraphics[width=0.8\linewidth]{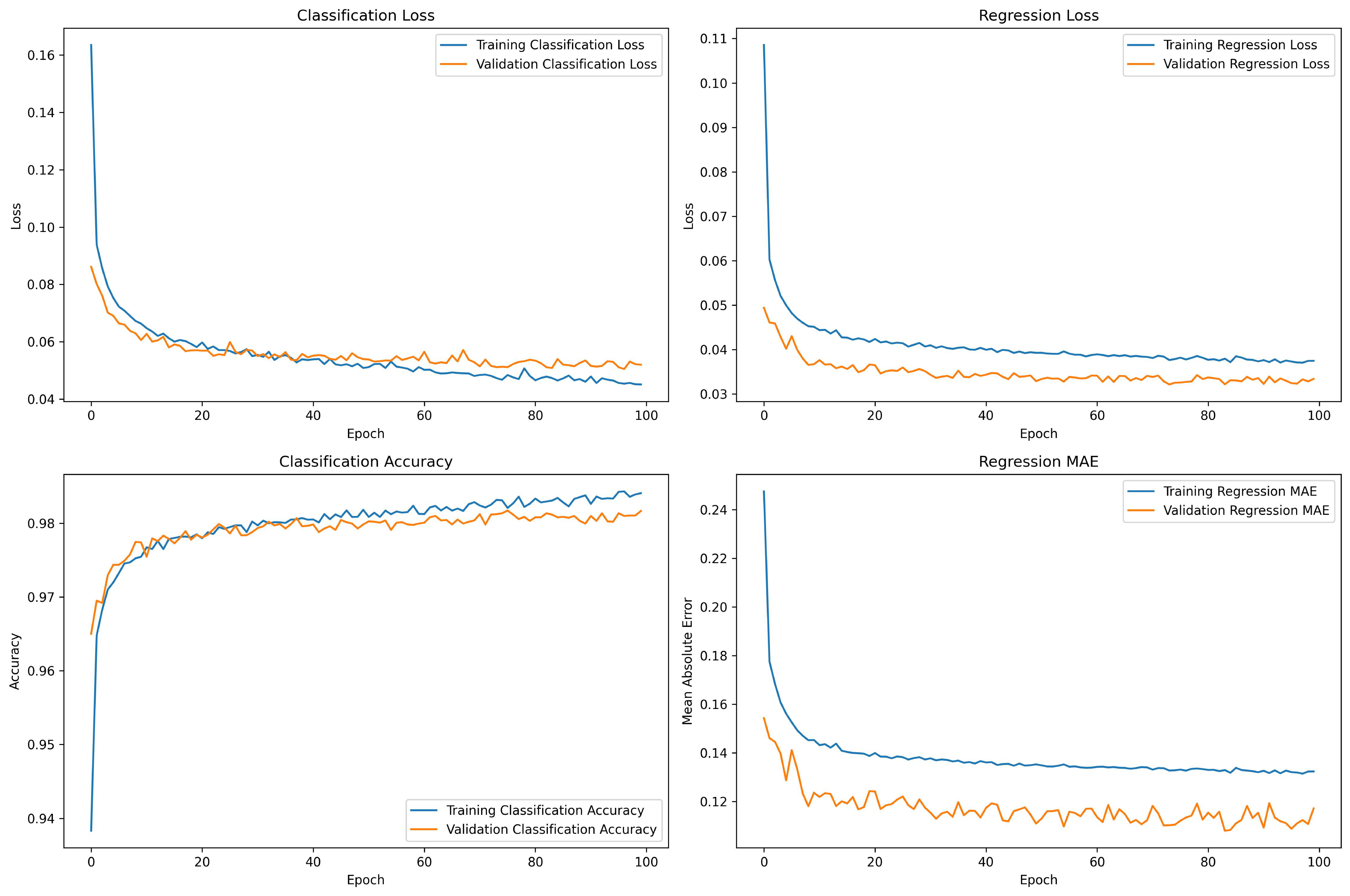}
    \caption{The upper left panel is the confusion matrix, and the upper right panel is the one-to-one correlation diagram. In the lower panel, we show the training loss information.}
    \label{fig:appendix_CNN}
\end{figure}

As shown in Figure \ref{fig:appendix_CNN}, the MLP achieves an overall classification accuracy of 98.16\%, with a GRB recall of 99.21\%, indicating that only a small number of GRBs are misclassified as background. For the regression task, the RMSE of the predicted spectral index is 0.1815, which is notably higher than that of the MLLM-based model. Moreover, among the 2,458 background observations, 2,455 yield positive regression outputs—a physically implausible result that does not occur in the MLLM-based predictions.

% \begin{thebibliography}{99}
\bibliographystyle{raa}
\bibliography{ms.bib}
% \end{thebibliography}

\label{lastpage}

\end{document}